\documentclass[a4paper,fleqn,titlepage,twocolumn,superscriptaddress]{revtex4-2}
\usepackage[utf8]{inputenc}
\usepackage{graphicx}
\usepackage{amsmath}
\usepackage{amssymb}
\usepackage{times}
\usepackage[utf8]{inputenc}
\usepackage{nicefrac}
\usepackage{color}
\usepackage{xcolor}
\definecolor{darkblue}{rgb}{0.0, 0.18, 0.65}
\usepackage{todonotes}
\usepackage{hyperref}
\hypersetup{
     colorlinks=true,
     linkcolor=darkblue,
     filecolor=darkblue,
     citecolor = darkblue,      
     urlcolor=darkblue}

\newcommand{\ignore}[1]{}

\newcommand{\U}[2]{\mathop{}\!\mathcal{U}_{#1}^{#2}}
\newcommand{\sle}[1]{\le^{#1}}
\definecolor{darkgreen}{rgb}{0.07, 0.53, 0.03}   
\begin{document}
\title{Order-disorder transition in the zero-temperature Ising model on random graphs}

\author{Armin Pournaki}
\affiliation{Max Planck Institute for Mathematics in the Sciences, Leipzig, Germany}
\affiliation{Laboratoire Lattice, CNRS \& ENS-PSL \& Université Sorbonne Nouvelle, Paris, France}
\affiliation{Sciences Po, médialab, Paris, France}
\author{Eckehard Olbrich}
\affiliation{Max Planck Institute for Mathematics in the Sciences, Leipzig, Germany}
\author{Sven Banisch}    
\affiliation{Karlsruhe Institute for Technology, Karlsruhe, Germany}
\author{Konstantin Klemm}
\affiliation{Institute for Cross-Disciplinary Physics and Complex Systems IFISC (UIB-CSIC), 07122 Palma de Mallorca, Spain}
\date{\today}

\begin{abstract}

The zero-temperature Ising model is known to reach a fully ordered ground state in sufficiently dense random graphs. In sparse random graphs, the dynamics gets absorbed in disordered local minima at magnetization close to zero. Here, we find that the non-equilibrium transition between the ordered and the disordered regime occurs at an average degree that slowly grows with the graph size. The system shows bistability: The distribution of the absolute magnetization in the reached absorbing state is bimodal, with peaks only at zero and unity. For a fixed system size, the average time to absorption behaves nonmonotonically as a function of average degree. The peak value of the average absorption time grows as a power law of the system size. These findings have relevance for community detection, opinion dynamics, and games on networks.
\end{abstract}

\maketitle

\section{Introduction}

The Ising model is a cornerstone of equilibrium statistical mechanics \cite{Ising:1925,Brush1967Ising}. Beyond its original scope of describing ferromagnetic phenomena, the model is a general reflection of discrete units' tendency to align their state with neighbors, e.g., agents' opinions in social systems \cite{Castellano2009social}. With temperature in a canonical ensemble playing the role of noise or deviations from the alignment tendency, the model at temperature zero is relevant as a noiseless base case. The zero-temperature limit of the equilibrium model, however, is not equivalent to the actual kinetics with temperature fixed at zero. While the former simply assigns all probability mass to the ground state configurations, the latter explicitly probes the energy landscape, especially local minima above the ground state energy, if the initial spins are drawn uniformly at random, which corresponds to an initialization at infinite temperature. The energy landscape, in turn, is generated by the underlying graph or interaction network.

We analyze the zero-temperature Ising model on random graphs \cite{gilbert1959random,erdHos1959random} for varying edge probability $p$ and network size $N$. This system is also known as the randomly dilute Curie-Weiss model (CW), which corresponds to a homogeneous CW in which the fixed interactions between all spin pairs are replaced by random ferromagnetic coupling [independent and identically distributed (i.i.d.) Bernoulli random variables with mean $p$] between any pair of spins. This change complexifies the energy landscape by introducing metastable states to which the system tends to converge at zero temperature \cite{Spirin2001,Svenson2001,Gheissari2018}. While similar behavior has been observed on other network topologies \cite{Das2005,Biswas2011,Khaleque2016} and update dynamics \cite{Castellano2005,Mukherjee2020}, we situate our findings at the gap between two previous results in the context of the randomly dilute CW: For sparse random graphs [$k \ll \log(N)$ \cite{Gheissari2018}, where the mean degree $k=p(N-1)$], the probability to reach the global minimum in which all spins are aligned (which we will refer to as \textit{consensus}) tends to 0 as $N \rightarrow \infty$ \cite{Haggstrom2002}. For dense random graphs ($k = O(N)$), the probability to reach consensus tends to 1 as $N \rightarrow \infty$ \cite{Gheissari2018}. All these results are asymptotic, for $N \rightarrow \infty$. Our contribution sheds light on the regime between \textit{finite} sparse and dense graphs, where the average degree $k \approx \log(N)$. Looking at final magnetization averaged over different realizations $\langle |m|_{\text{final}} \rangle$, we observe a transition from disorder ($\langle |m|_{\text{final}} \rangle \approx 0$) to consensus ($\langle |m|_{\text{final}} \rangle \approx 1$) as we move from sparse to dense random graphs by increasing the mean degree. The dynamical runs that do not reach consensus get trapped in local minima. As the networks densify, the probability to reach a local minimum decreases continuously, facilitating the system to reach consensus. With growing network sizes $N$, we observe that the transition from disorder to consensus shifts to higher values of $k$. 

\section{Dynamics}
We investigate the ferromagnetic Ising model at zero temperature on random graphs \cite{gilbert1959random,erdHos1959random}. Consider a graph $G=(V,E)$, where $V=\{1,\dots,n\}$ is the set of vertices and $E$ the set of edges. Every vertex $i \in V$ has a binary state variable $s_i \in \{-1,+1\}$. A spin configuration $s=(s_1,\dots,s_n)$ is assigned the energy
\begin{equation} \label{eq:hamiltonian}
H(s) = -\sum_{\{i,j\} \in E} s_i s_j~.
\end{equation}

We consider zero-temperature Metropolis dynamics with initial conditions $s(0) \in \{-1,+1\}^V$ drawn uniformly at random. In every microstep, a random vertex $i$ is chosen. The state of $i$ is flipped, $s_i \rightarrow - s_i$, if the resulting configuration has lower than or equal energy as $s$ itself [cf.\ Eq.~(\ref{eq:hamiltonian}]. Time is updated as $t \rightarrow t + 1/n$ in every microstep.
Microsteps are iterated until the dynamics reaches a local minimum $s^\text{final}$, defined as a configuration from where configurations with strictly lower energy are not reachable by the Metropolis dynamics. Since the drawn graphs contain nodes with an even number of neighbors, adjacent configurations with equal energies occur, which in the literature are referred to as \textit{blinker states} \cite{Olejarz2011}. Thus a given configuration cannot be identified as a local minimum based on its neighboring configurations (with respect to single-spin flips) alone. See the Appendix for details of our method for detection of local minima. 

For each run, we consider the magnetization of the local minimum reached:
\begin{equation}
m_\text{final} = \frac{1}{N} \sum_{i \in V} s_i^\text{final}
\end{equation}
For the absolute value of the final magnetization, we compute the first moment $\langle |m_\text{final}| \rangle$ and the variance (centered second moment) $\sigma^2=\langle m^2_\text{final} \rangle - \langle |m_\text{final}| \rangle^2$
where $\langle \cdot \rangle$ denotes averaging over realizations. We also record the time $T$ until a dynamical run first reaches a local minimum and consider the average $\langle T \rangle$ over realizations.

For given values of $N$ and average degree $k$, $100$ random graphs are generated and $1000$ independent runs are performed on each of the graphs, resulting in $10^5$ runs per parameter choice. A graph $G$ is generated as follows. Draw a random graph from the ensemble $\mathcal{G}(N,p)$ with edge probability $p=k/(N-1)$; take as $G$ the connected component with a maximum number of nodes $n$ and disregard all other connected components. 

\section{Results}

\begin{figure}
    \centering
    \includegraphics[width=.49\textwidth]{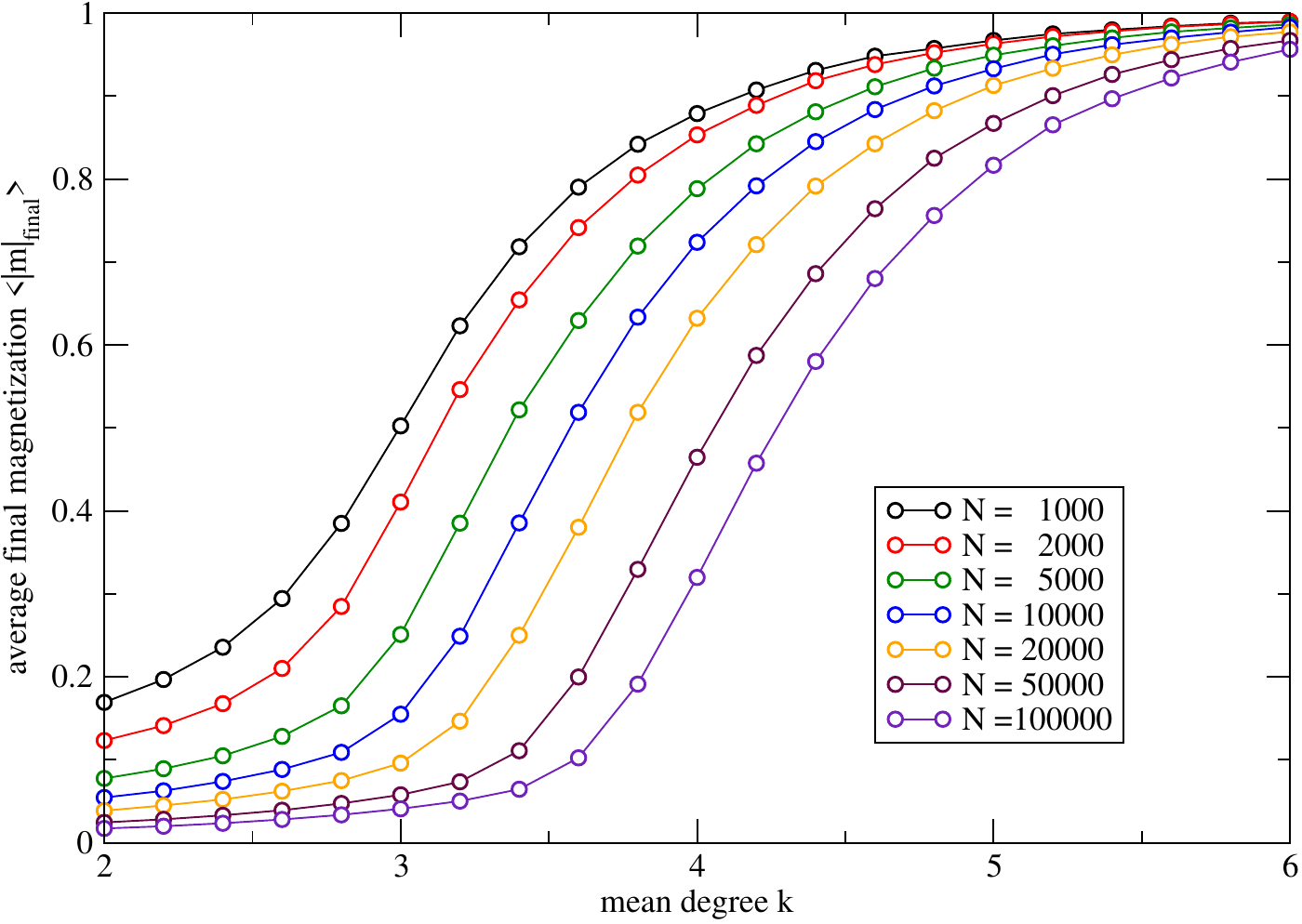}
    \caption{Average final magnetization $ \langle |m|_{\text{final}} \rangle$ as a function of the mean degree $k=p(N-1)$ for different network sizes, increasing from the top curve to the bottom curve. We observe a transition from disorder ($|m|_{\text{final}} \approx 0$) to consensus ($|m|_{\text{final}} \approx 1$). With growing network sizes, the transition shifts to higher values of $k$.}
    \label{fig:discmag_0}
\end{figure}
\begin{figure}
    \centering
    \includegraphics[width=.499\textwidth]{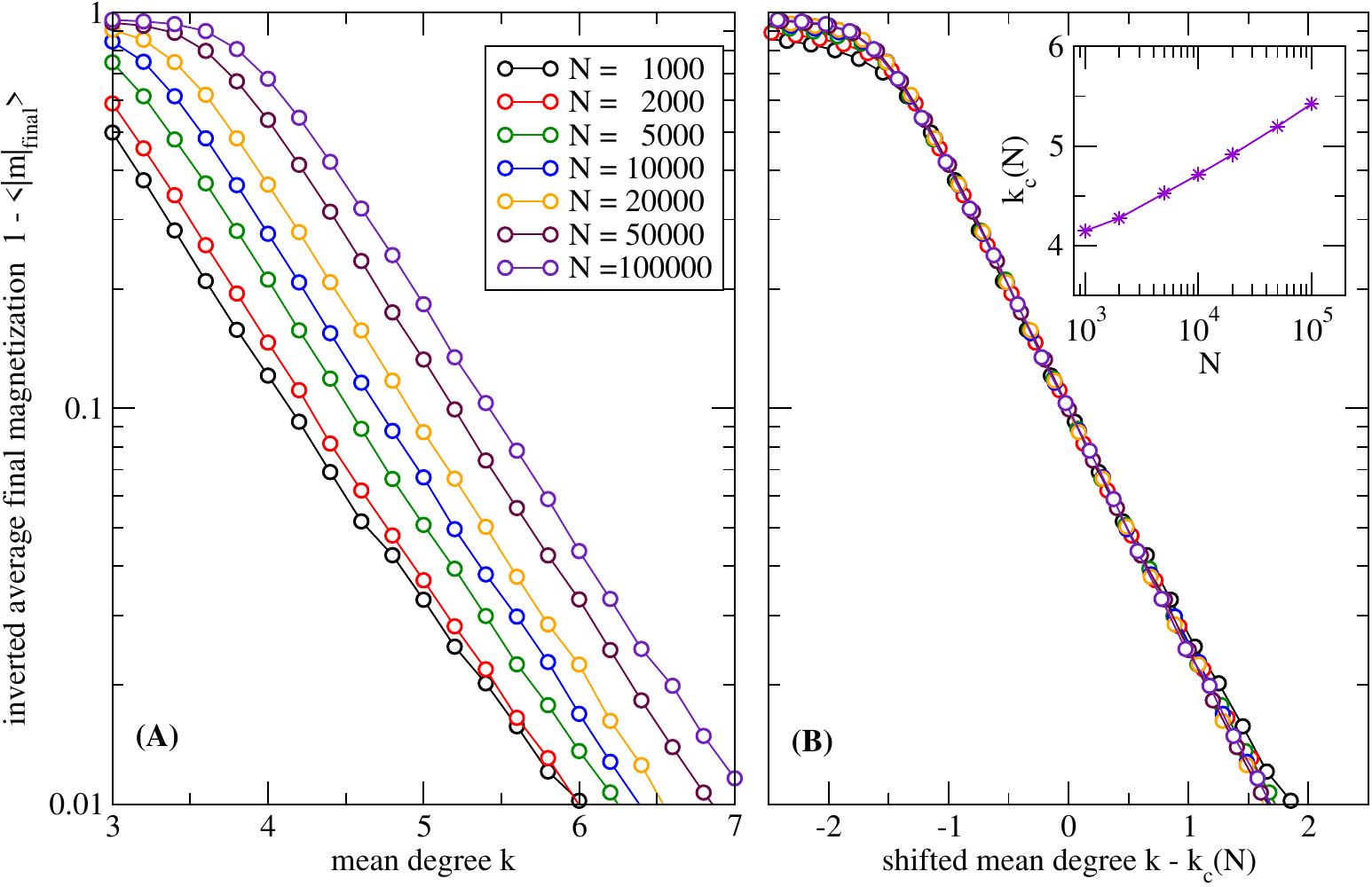}
    \caption{Inverted average final magnetization $1- \langle |m|_{\text{final}} \rangle$ as a function of the mean degree $k$ for different network sizes, decreasing from the top curve to the bottom curve. \textbf{A} shows that the inverted average final magnetization decreases exponentially for $k > 3$. In \textbf{B}, we shift the mean degree [$k \rightarrow k-k_C(N)$] such that the curves overlap to reveal the $N$-dependency of the transition point shown in the inset. Using $p=k/N$ as the control parameter, we would get a critical $p_c \approx \log(N)/N$, which tends to 0 as $N$ tends to infinity.}
    \label{fig:discmag_1}
\end{figure}
\begin{figure}
    \centering
    \includegraphics[width=.49\textwidth]{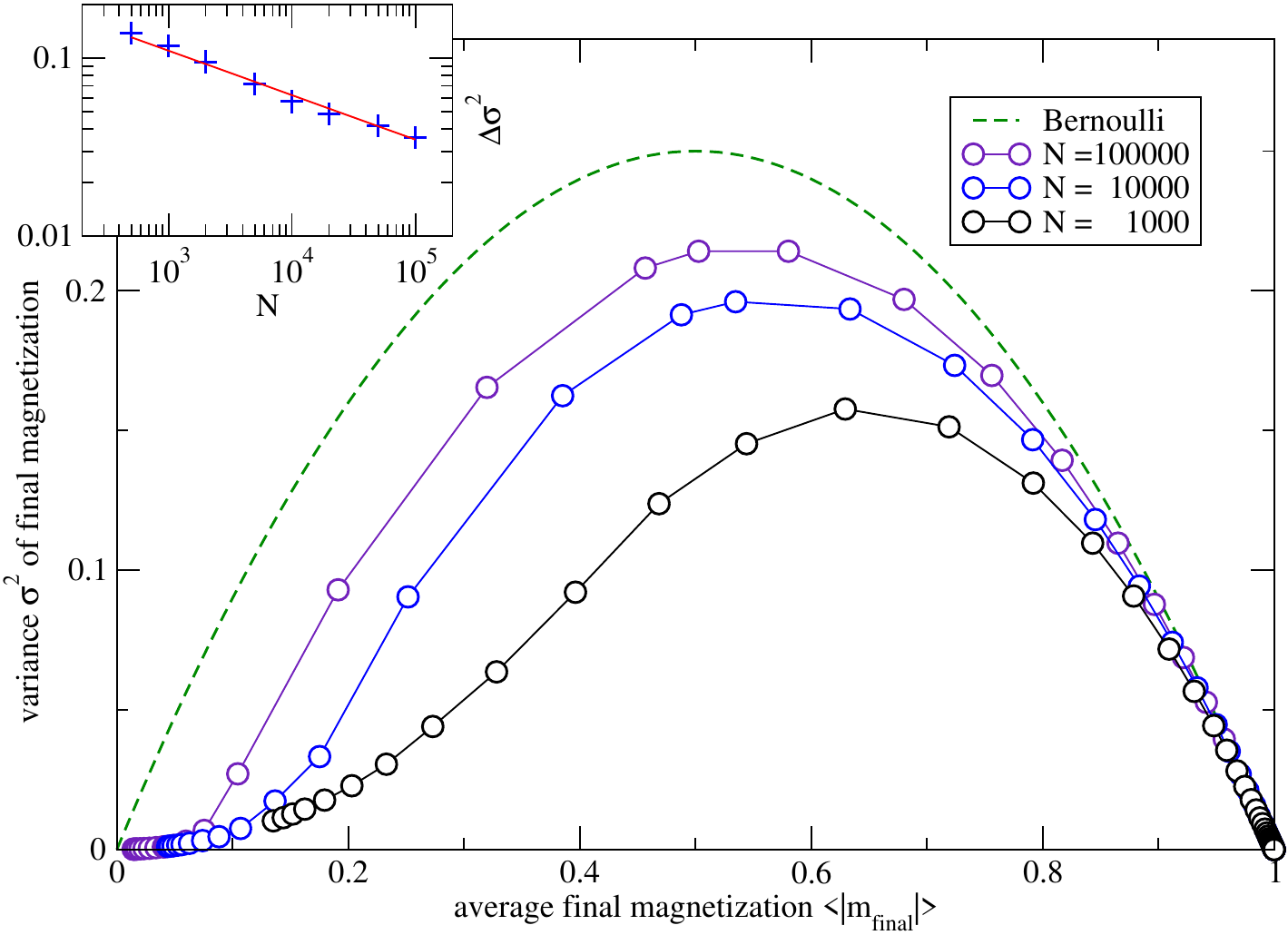}
    \caption{Variance vs average final magnetization for different network sizes, decreasing from the top curve to the bottom curve. The dashed line indicates the maximally possible variance at the given mean value, as obtained for a distribution with the given mean but restricted to values at 0 and 1. The inset shows the system size dependence of $\Delta\sigma^2 = 0.25-\sigma^2$ evaluated at $\langle|m|\rangle=0.5$ ($+$ symbols) and a straight line with slope $-0.25$.}
    \label{fig:discvar_0}
\end{figure}
\begin{figure}
    \includegraphics[width=.35\textwidth]{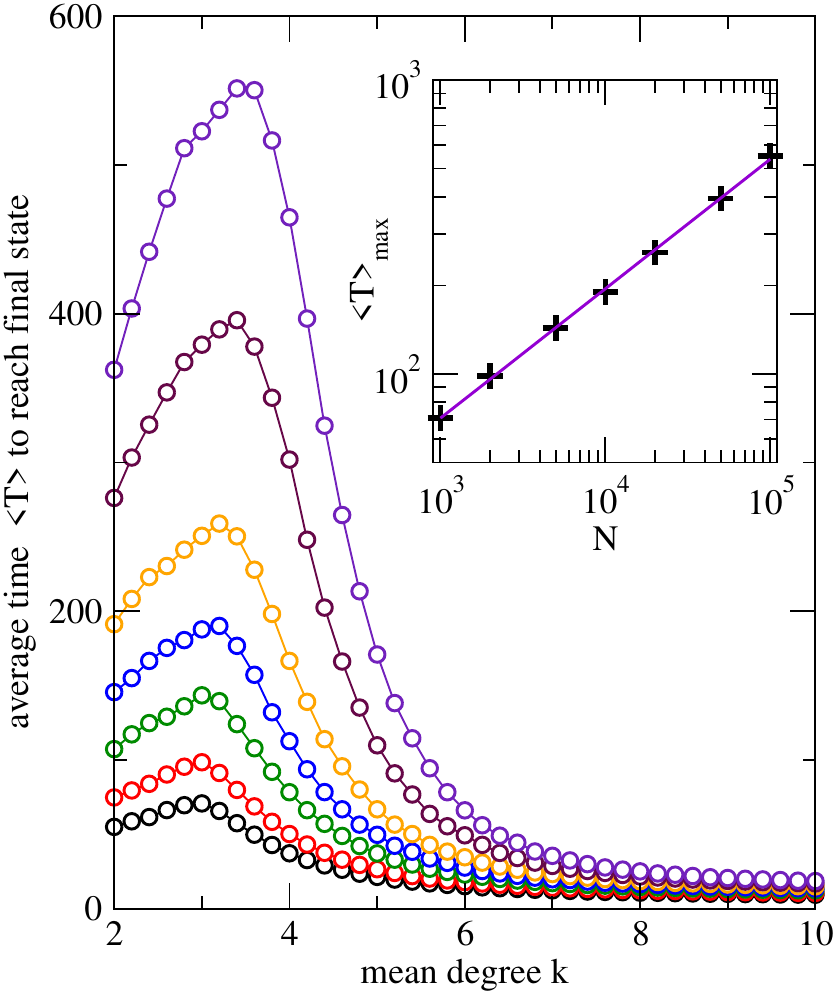}
    \caption{Average time to reach the final state as a function of the mean degree $k$ for different network sizes, decreasing from the top curve to the bottom curve. The peak, designating the network parameters that take the longest time to reach the final state, is close to $k=\log(N)$. With growing network sizes, this peak again shifts to higher $k$. For each curve of the main panel, the inset shows the peak value $T_\text{max}$  as a function of $N$ ($+$ symbols); the solid line is the result of a power-law fit yielding an exponent $0.440(5)$, and a correlation coefficient 0.9996. Colors designate network sizes and are the same as in Fig.~\ref{fig:discmag_0}.}
    \label{fig:disctime_0}
\end{figure}
\begin{figure}
    \includegraphics[width=.49\textwidth]{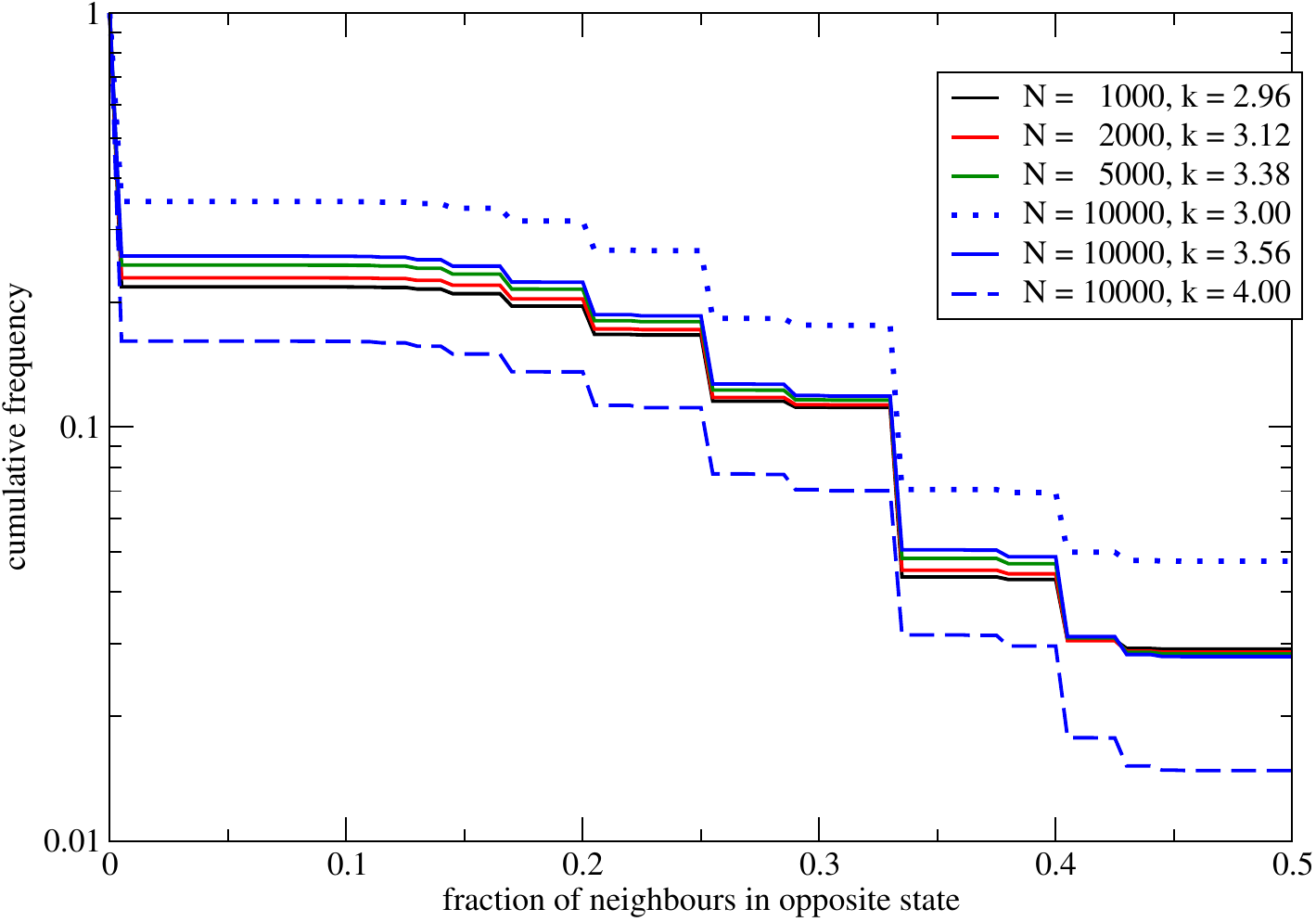}
    \caption{Cumulative histograms of the fraction of opposing neighbors in the final state, taken over all nodes of degree $ k>0$ in 10000 independent runs for a given $N$ and $k$. Network sizes are increasing from the top solid curve to the bottom solid curve. For the solid curves, $k$ is chosen such that the final average magnetization $ \langle |m|_{\text{final}} \rangle = 0.5$. The step shape of histogram is due to the fact that these fractions are the result of divisions by small integers.}
    \label{fig:inter_0}
\end{figure}

Fig.~\ref{fig:discmag_0} shows the average absolute value of the magnetization of the system's final state $\langle|m|_{\text{final}}\rangle$ as a function of the mean degree $k=p(N-1)$ for different network sizes. As we increase the mean degree, we observe a smooth transition from disorder ($|m|_{\text{final}} \approx 0$) to consensus ($|m|_{\text{final}} \approx 1$). Simulating the dynamics on different network sizes (from $N=10^3$ to $N=10^5$), we observe that with growing network size, this transition happens at higher mean degrees $k$. 
Looking at the inverted average final magnetization $1 - \langle|m|_{\text{final}}\rangle$ in Fig.~\ref{fig:discmag_1}A, we observe that it decays exponentially for sufficiently large $k$.
To quantify the influence of the network size, we shift the magnetization curves: We define $k_c(N)$ as the value of $k$ where $1-|m| = 0.1$. Fig.~\ref{fig:discmag_1}B shows the data collapse of the magnetization plotted as a function of $k - k_c$, the inset showing the values $k_c(N)$. We see that the critical degree $k_c(N)$ grows at least as $\log(N)$, which lies between the growth regimes of sparse and dense graphs: The former are characterized by $k \ll \log(N)$ \cite{Gheissari2018}, while the latter exhibit $k = O(N)$. Therefore, the transition from disorder to order in the randomly dilute CW model happens in a parameter region between sparse and dense graphs.

To further quantify this transition, we examine the distribution of the magnetization for different graph realizations and initial conditions in Figure~\ref{fig:discvar_0}. For increasing network size $N$, this distribution approaches a Bernoulli distribution where $|m|_\text{final}$ takes values 0 and 1 only. For  a mean value of $0.5$, the Bernoulli distribution has a variance $\sigma^2=0.25$. As shown in the inset of Figure~\ref{fig:discvar_0}, the difference $\Delta\sigma^2$ between observed variance and the theoretical maximum decreases algebraically as we increase $N$ while keeping $ \langle |m|_{\text{final}} \rangle = 0.5$. If this trend extrapolates to the limit of large $N$, the system either reaches consensus or is trapped in a local minimum with $|m|_\text{final} \approx 0$.

Fig.~\ref{fig:disctime_0} shows the average time $\langle T \rangle$ to reach the final state as a function of the average degree $k$ for various system sizes. It behaves non-monotonically, as there is a peak value of $\langle T \rangle$ for a given system size $N$ which increases as a power law with $N$ (see the inset of Fig.~\ref{fig:disctime_0}). We may compare this behavior to the case of complete graphs, where the average time to reach the final state is given by $\langle T \rangle \sim \log (N)$.

In order to characterize the final states of the system, we compute for each node the fraction of neighbors in the opposite state. This quantity is given by 
\begin{equation}
    \pi_s(i) = 1-\frac{1}{k_i} \sum_{j \in N(i)} \delta_{s_i,s_j}
\end{equation}
where $N(i)$ designates the neighborhood of $i$, $k_i$ its degree, and $\delta_{s_i,s_j}=1$ if $s_i=s_j$. The distribution of $\pi_s(i)$ gives insight into the structure of the local minima. If all values are strictly below $1/2$, this means that the associated state is strictly stable because no individual nodes would flip spins under the dynamics. 
Notice that in a consensual state $\pi_s(i) = 0$ for all nodes. On the other hand, values $\pi_s(i) > 1/2$ indicate that the node would switch under the dynamics, rendering an associated spin profile $s$ unstable. One important question concerns the existence of local minima that contain blinkers, that is, nodes with an equal number of aligned and unaligned neighbors such that their spin flip does not change the global energy of the system \cite{Baek2012}. These nodes are characterized by $\pi_s(i) = 1/2$.

Fig.~\ref{fig:inter_0} shows the cumulative histogram of these fractions $\pi_s(i)$ for different network sizes and values of mean degree $k$ such that $ \langle |m|_{\text{final}} \rangle = 0.5$, where roughly half of the runs land in consensus and the other half in a local minimum. We see that the majority of nodes have fully homogeneous neighborhoods and that the probability to observe larger $\pi_s(i)$ decreases in a sequence of steps. 
We observe a considerable proportion of ``indifferent'' agents (blinkers), who have the same amount of neighbors in each state at the end of the dynamics. This indicates the existence/prevalence of local minima defined by a series of neighboring states $s$ that can be reached by flipping single nodes' spins.




Finally, in order to check the robustness of our results, we ran the dynamics on graphs with continuous-valued positive coupling  strengths, as well as two modified dynamics: Glauber dynamics, where equal-energy spin flips happen only with probability $1/2$, and another type of dynamics where spin flips happen only if they \textit{strictly} decrease the energy. All three scenarios present qualitatively the same transition as seen in Fig.~\ref{fig:discmag_0} at a critical mean degree increasing with system size as well.

\section{Discussion}
We have studied the transition from disorder to order in zero-temperature dynamics on the randomly dilute Curie-Weiss model as we move from sparse to dense random graphs by increasing the mean degree. The transition depends on the graph size $N$, which shifts the critical degree to higher values of $k$. 
In the transition region realizations can be trapped in a plethora of local minima with zero magnetization ($|m|_{\text{final}} \approx 0$).

We may describe the persistence of these local minima in the context of community structure in networks. As we have seen for certain configurations of $N$ and $k$, the system consistently converges to metastable states, which raises the question whether the graph partitions given by these states can be related to partitions in the sense of community detection, such as the ones gained by modularity maximization \cite{Newman2004a}. The intuition is that \textit{communities} correspond to clusters of nodes that are more strongly connected within their cluster than across the network. Following this approach, a given partition $b$ of a graph $G$ is evaluated using the following function,
\begin{equation}
  \label{eq:mod1}
  Q(A,b) = \frac{1}{2m}~\sum_{i,j}~(A_{ij}-\frac{k_ik_j}{2m})~\delta_{b_i,b_j}
\end{equation}
where the entry of the adjacency matrix $A_{ij} = 1$ if there exists an edge between nodes $i$ and $j$, $0$ otherwise. The normalization factor $2m=\sum_{i,j}A_{ij}$ makes the measure comparable across network sizes. The community index of node $i$ is denoted by $b_i$ and $\delta_{b_i,b_j} = 1$ if $i$ and $j$ are in the same community, $0$ otherwise. 
The term $k_ik_j/2m$ implements the configuration model \cite{Chung2002,fosdick2018configuring} as the null model to which the real network structure is compared.
The modularity $Q^*(A)$ of a graph is then defined as the maximum modularity of all its possible partitions:
\begin{equation}
  \label{eq:mod2}
  Q^*(A) = \max_b~Q(A,b) 
\end{equation}
%
Previous works have shown that for random graphs, $\lim_{k \to \infty}Q^* \to 0$ \cite{Guimera2004,McDiarmid2020}. The metastable traps the system converges to in the case of the randomly dilute Curie-Weiss model discussed here therefore correspond to partitions that are not uncovered by modularity maximization methods.

Beyond the energy landscape of the Ising model, the question addressed by this paper is also relevant in the context of opinion dynamics on social graphs \cite{Castellano2009social}. Our results show that even in random graphs there is a certain regime of graph connectivity | in between sparse and dense graphs | in which non-consensual opinion profiles 
can be a stable outcome on connected components. This is remarkable because random graphs are characterized by the absence of group structures and network segregation which were assumed to be the 
driving forces behind polarization dynamics in many opinion dynamics models \cite{Friedkin2011social,Maes2013differentiation,Banisch2019opinion}. 

Our results are also interesting in the context of coordination games played over a social network \cite{Jackson2015games} because the observed zero-temperature dynamics in the randomly dilute CW model can be considered as a best response update in a symmetric coordination game where the payoff of agent $i$ is $\sum_{j \in N(i)} s_i s_j$, which is the negative local energy of node $i$. In this case, one is interested in the set of possible Nash equilibria defined as network configurations ($s$) in which no agent alone is better off by changing their action. This definition hence corresponds to how local minima have been defined in the present paper. In games on networks, the existence of non-consensual equilibria is captured by the notion of cohesive sets, which partition the set of nodes in a network in two such that every agent has more connections to its own subset than to the other \cite{Morris2000contagion}. The existence of local minima in the randomly dilute CW model therefore proves that non-consensual outcomes in symmetric coordination games can be stable on random graphs. An interesting follow-up question would be to relate the number of local minima in the energy landscape to the number of mutually disjoint cohesive sets of the network.

%
\section{Acknowledgements}
We thank Felix Gaisbauer for helpful comments and suggestions. A.P. was funded in part by the French government under management of Agence Nationale de la Recherche as part of the "Investissements d'avenir" program, reference ANR-19-P3IA-0001 (PRAIRIE 3IA Institute). E.O. and S.B. acknowledge support from the European Union’s Horizon 2020 research and innovation programme under grant agreement No 732942 (ODYCCEUS). K.K. acknowledges support from Project No. PID2021-122256NB-C22 funded by MCIN/AEI/10.13039/501100011033 / FEDER, UE.

\bibliographystyle{apsrev4-2}
\bibliography{bib}
\appendix
\section{Identification of local minima under neutrality} 

For a configuration $s = (s_1,s_2,\dots,s_n) \in S^V$, node $i\in V$ and $\sigma \in S$, we write $\U{i}{\sigma} s$ as the configuration having $\sigma$ as the entry at node $i$ and being $s_j$ for all other indices $j\neq i$. This way we describe updating the spin at node $i$ with direction $\sigma$. The change in energy caused by the update is
\begin{equation} \label{eq:Hdiff}
H(\U{i}{\sigma} s) - H(s) = - (\sigma- s_i) \sum_{j \in \Gamma(i)} s_j
\end{equation}
with $\Gamma(i):= \{j \in V : \{i,j\}\in E\}$ being the neighbourhood of node $i$.

For configurations $r,s \in S^V$ and $\sigma \in S$, we say that $r$ is a $\sigma$-minor of $s$, in symbols $r \sle{\sigma} s$, if $s_i \in \{r_i,\sigma\}$ for all $i \in V$.

\underline{Lemma 1:} Let $r, s \in S^V$ and $\sigma \in S$ so that $r \sle{\sigma} s$. Furthermore consider $i \in V$ with $s_i \neq \sigma$. Then 
\begin{equation}
H(\U{i}{\sigma} r) - H(r) \ge H(\U{i}{\sigma} s) - H(s)~.
\end{equation}

\underline{Proof:} 
Due to $s_i \neq \sigma$ and $r \sle{\sigma} s$, we find $(\sigma-r_i) = (\sigma-s_i) = 2\sigma$. Therefore equation (\ref{eq:Hdiff}) implies 
$H(\U{i}{\sigma} r) - H(r) - H(\U{i}{\sigma} s) + H(s) = - 2 \sigma \sum_{j \in \Gamma(i)} (r_j - s_j)$. 
Since $r$ is a $\sigma$-minor of $s$, we have $\sigma r_j \le \sigma s_j$ for all $j \in V$, which implies $-2 \sigma \sum_{j \in \Gamma(i)} (r_j - s_j) \ge 0$ and completes the proof. $\square$

A {\em walk} (of length $l$) is a sequence $s(0),s(1),s(2),\dots,s(l)$ of configurations in $[q]^V$ where for each $k \in [l]$ there are $i\in V$ and $\sigma \in [q]$ so that $s(k) = \U{i}{\sigma} s(k-1)$. For fixed $\sigma \in [q]$, a walk $s(0),s(1),s(2),\dots,s(l)$ is called {\em $\sigma$-homogeneous} if for each $k \in [l]$ there is $i \in V$ so that $s(k) = \U{i}{\sigma} s(k-1)$.

A walk $s(0),s(1),s(2),\dots,s(l)$ is {\em adaptive} if $H(s(k))\le H(s(k-1))$ for all $k \in [l]$. The adaptive walk $s(0),s(1),s(2),\dots,s(l)$ is {\em escaping} if $H(s(k))<H(s(k-1))$ if and only if $k=l$. A configuration $s \in S^V$ is called {\em local minimum} if all adaptive walks starting in $s$ are not escaping. By the following lemmata we establish the existence of a homogeneous escaping walk from any configuration not being a local minimum.  

\underline{Lemma 2:} Consider a configuration $s(0) \in S^V$, and suppose there is an escaping walk $s(0),s(1),\dots,s(l)$. Then there are $\sigma \in S$, $l^\prime \le l$ and a $\sigma$-homogeneous escaping walk $r(0),r(1),\dots,r(l^\prime)$ with $r(0)=s(0)$.

\underline{Proof:} Choose $s(0),s(1),\dots,s(l)$ as an escaping walk of minimum length $l$. Find $i \in V$ and $\sigma \in V$ so that $s(l)=\U{i}{\sigma} s(l-1)$. We make the assumption (to be led to contradiction) that the walk $s(0),s(1),\dots,s(l)$ is not $\sigma$-homogeneous. Then there are $k \in [l-1]$ and $j \in V$ so that $s_j(k-1)=\sigma$ and $s_j(k)\neq \sigma$. Choose such $k$ and $j$ so that $k$ is maximal. Define a walk $r(0),r(1),\dots,r(l-1)$ by setting $r(m) = \U{j}{\sigma} s(m+1)$ for
$k\le m < l$ and $r(m)=s(m)$ otherwise. By construction for each $m \in \{k,\dots,l-1\}$, $s(m+1)$ is a $\sigma$-minor of $r(m)$ implying
$H(\U{i}{\sigma} r(m)) - H(r(m)) \le H(\U{i}{\sigma} s(m+1)) - H(s(m+1))$.
Therefore $H(r(l-1))\le H(s(l))$, so there exists an escaping walk from $s(0)$ strictly shorter than $l$, incompatible with the choice of $s(0),s(1),\dots,s(l)$ as having minimum length. The assumption that this escaping walk is not $\sigma$-homogeneous thus leads to a contradiction, completing the proof. $\square$

\underline{Lemma 3:} Consider $\sigma \in [q]$ and a configuration $s(0) \in [q]^V$, and suppose there is a $\sigma$-homogeneous escaping walk $s(0),s(1),\dots,s(l)$. Furthermore suppose there is $i \in V$ with $r(0) = \U{i}{\sigma} s(0)$ having $H(r(0))=H(s(0))$. Then there is a $\sigma$-homogeneous escaping walk $r(0),r(1),\dots,r(m)$.

\underline{Proof:} Construct the walk by doing the same updates as in the given walk. Since all updates are $\sigma$-updates, the energy along the constructed walk is less or equal to that of the given one, i.e. $H(r(k))\le H(s(k))$ for all $k \in [l]$. $\square$

\underline{Algorithm:} Given a configuration $s \in [q]^V$, we use the following method to decide if $s$ is a local minimum. An outer loop runs over all spin directions $\sigma \in S$. For each value of $\sigma$, we initialize $r=s$ and run the following inner loop. (i) If there is a node $i$ with $r_i \neq \sigma$ and $H(\U{i}{\sigma} r) \le H(s)$, update $r \leftarrow \U{i}{\sigma} r$; otherwise leave the inner loop. (ii) If $H(r)<H(s)$ terminate with result {\em not a local minimum}. (iii) resume at (i).

If the execution ends without result {\em not a local minimum} (step (ii)), then the result is that $s$ is a {\em local minimum}.

\underline{Proof of correctness:} Suppose first that $s$ is not a local minimum so $s$ has an escaping path. By Lemma 2, find $\sigma \in S$ and a $\sigma$-homogeneous escaping walk from $s$. By Lemma 3, also all configurations reachable from $s$ by a $\sigma$-homogeneous walk have a $\sigma$-homogeneous escaping walk. Thus, when the inner loop with the right $\sigma$ is performed, such an escaping walk to a lower energy configuration $r$ will be found, terminating with result {\em not a local minimum}. Conversely, suppose $s$ is a local minimum so there is not an adaptive walk leading to a configuration with energy below $H(s)$. Since only non-increasing updates lead from $s$ to any configuration $r$ encountered, $H(r) =H(s)$. The condition at (ii) is never fulfilled so the algorithm will return {\em local minimum}. 

\end{document}